\documentclass[aps,pre,reprint,pdftex]{revtex4-1}
\usepackage{amsmath,amssymb,amsthm}
\usepackage[dvips]{graphicx,color}
\usepackage[pdftex,colorlinks]{hyperref}

\newcommand{\bra}[1]{\left\langle #1 \right|}
\newcommand{\ket}[1]{\left| #1 \right\rangle}

\begin{document}

\title{The Accuracy of Perturbative Master Equations}
\author{C. H. Fleming, N. I. Cummings}
\affiliation{Joint Quantum Institute and Department of Physics, University of Maryland, College Park, Maryland 20742}
\date{January 12, 2011}

\begin{abstract}
We consider open quantum systems with dynamics described by master equations that have perturbative expansions in the system-environment interaction.
We show that, contrary to intuition, full-time solutions of order-$2n$ accuracy require an order-$\left(2n\!+\!2\right)$ master equation.
We give two examples of such inaccuracies in the solutions to an order-$2n$ master equation:
order-$2n$ inaccuracies in the steady state of the system and order-$2n$ positivity violations,
and we show how these arise in a specific example for which exact solutions are available.
This result has a wide-ranging impact on the validity of coupling (or friction) sensitive results derived from second-order convolutionless, Nakajima-Zwanzig, Redfield, and Born-Markov master equations.
\end{abstract}

\maketitle

%\tableofcontents

%%%%%%%%%%%
\section{Introduction\label{sec:intro}}
An open quantum system is a quantum system that interacts with some environment whose degrees of freedom have been coarse grained over (i.e., traced out), and its dynamics are described by a master equation governing the reduced density matrix $\boldsymbol{\rho}$.
Exact master equations for the stochastic dynamics of open quantum systems are, in general, out of reach.
However, arbitrary-order perturbative master equations (in the system-environment interaction) can be derived in a variety of different ways \cite{Kampen97,Breuer01,Strunz04}
and find application in many branches of physics and chemistry \cite{Pollard97,Carmichael99,Breuer02,Kampen07}.
In the time-local representation (also called the convolutionless or Markovian representation), the dynamics of the reduced density matrix of the system can be expressed with a quantum Liouville equation
\begin{equation}
\frac{d}{dt} \boldsymbol{\rho}(t) = \boldsymbol{\mathcal{L}}(t) \, \boldsymbol{\rho}(t) \, \label{eq:TCME},
\end{equation}
where, despite the apparent time-local form, non-Markovian behavior may be encapsulated in the time dependence of the Liouvillian $\boldsymbol{\mathcal{L}}(t)$.
As a perturbative approximation, $\boldsymbol{\mathcal{L}}(t)$ is expanded in powers of the system-environment interaction, $c$, and truncated to some order.

% NC: Are all these assumptions necessary at this point, or are we imposing them before they're really necessary?

We will consider perturbative master equations where the Liouvillian $\boldsymbol{\mathcal{L}}(t)$ is time independent at zeroth order and asymptotically constant for late times.
We will assume that the perturbative expansion of $\boldsymbol{\mathcal{L}}(t)$ is in even powers of the coupling,
because as we explain in the next section, this can naturally arise from a microscopic derivation of the open-system dynamics.
The expansion of $\boldsymbol{\mathcal{L}}$(t) will then take the form 
\begin{align}
\boldsymbol{\mathcal{L}}(t) &= \sum_{n=0}^\infty  \boldsymbol{\mathcal{L}}_{[2n]}(t) \, ,\label{eq:PerturbExp} \\
\boldsymbol{\mathcal{L}}_{[0]}(t) \, \boldsymbol{\rho} &\equiv \left[ -\imath \, \mathbf{H} , \boldsymbol{\rho} \right] \, ,
\end{align}
where $ \boldsymbol{\mathcal{L}}_{[2n]}(t) = \mathcal{O}(c^{2n})$ and to zeroth order the system is driven in a unitary manner by its Hamiltonian $\mathbf{H}$.

The most well-known perturbative master equation is the second-order master equation, as it can be equivalent to the Redfield and Born-Markov master equations.
This is partly due to the fact that in the Markovian limit, the second-order master equation is exact.
But equivalence with the previous approximate master equations does not carry to fourth order and there perturbation theory is strictly superior.
One might easily assume that solving the second-order master equation defined by the Liouvillian  $\boldsymbol{\mathcal{L}}_{[0]}+ \boldsymbol{\mathcal{L}}_{[2]}$ would yield a solution that would match the exact solution to the exact master equation up to second order, having error terms of order $\mathcal{O}(c^{4})$;
however we will show that in general they will differ by second-order terms, so that one can only say they are in perturbative agreement at zeroth order.  

One very significant implication of these facts is for positivity.
Not being exact, nor generally of Lindblad form \cite{Lindblad76,Gorini76},
a perturbative master equation is not guaranteed to yield a dynamical map with exact complete positivity. 
Solutions can and should be completely positive to the relevant perturbative order,
and as we show in this work that order is not what one might naively expect.
Solutions to the second-order master equation can violate positivity by an amount that is $\mathcal{O}(c^{2})$.
We show that to find solutions good to second-order, canonical perturbation theory generally demands the fourth-order Liouvillian.

In our penultimate section, we use the example of quantum Brownian motion to show how naive usage of the second-order master equation leads to an underestimation of the second-order position uncertainty of a damped quantum oscillator (such as a nanomechanical resonator).
We would refer the reader to this section to see an immediate physical implication of the accuracy loss we describe in this work.

Finally, in our core proof we employ the formalism of canonical perturbation theory as applied to master equations \cite{QOS}.
If the reader is interested in seeing a specific example problem worked out in its entirety with these techniques,
we would refer the reader to Ref.~\cite{Dipole} which considered neutral atoms held fixed in the electromagnetic field.
There it was found that the second-order master equation is completely inadequate in assessing sudden death of entanglement at low temperature.
It was also in this work that we noticed the relation between second-order positivity loss and low temperature, which will be discussed later.

%%%%%%%%%%%%
\section{Master Equations from a Microscopic Model\label{sec:micro}}

In a system derived from a microscopic model, the coarse-grained environment can act as a source of noise, dissipation, and decoherence;
thus its influence provides a model of dissipative quantum mechanics more general than Markovian (white-noise) models, which can be constructed phenomenologically.
Such microscopic models can still lead to a stationary Liouvillian in the late-time limit, even when the dynamics they describe are non-Markovian.

Given a stationary system Hamiltonian and stationary bath correlations,
Gaussian noise distributionals (e.g. noise generated via linear coupling to an environment of harmonic oscillators) \emph{may} allow the master equation to have a stationary late-time limit \cite{QOS}
\begin{align}
\boldsymbol{\mathcal{L}}(\infty) &= \lim_{t \to \infty} \boldsymbol{\mathcal{L}}(t) \, , \label{eq:Linf}
\end{align}
so that the late-time and weak-coupling limits commute; otherwise perturbation theory cannot be used for long durations of time.
Gaussian noise processes are categorized by their second-order noise correlation, 
and whether or not the master equation will have a stationary limit is dependent upon how localized this noise correlation is. 
Well-localized noise correlations (e.g. Gaussian or exponential) can lead to a very well-behaved master equation,
whereas long-ranged noise correlations (e.g. Cauchy) can produce a more pathological master equation which cannot be accurately analyzed in a perturbative fashion.
Exact examples of this phenomena are given in Ref.~\cite{QBM} in the context of quantum Brownian motion with Ohmic and sub-Ohmic couplings.
Moreover, the exact solutions $\boldsymbol{\rho}(t)$ can be very well-behaved even if $\boldsymbol{\mathcal{L}}(t)$ is not.
Markovian representations (and, more generally, effective equations of motion) are not always suitable.

%%%%%%%%%%%
\section{Indeterminacy of Solutions}
Before determining what the appropriate level of accuracy is for the solutions of perturbative master equations,
we will first demonstrate that there is issue with the naive expectation of order-$2n$ accuracy.
This argument is a generalization of one found in Ref.~\cite{Mori08}, where the discrepancy was noticed for the second-order equilibrium state.
Let $\boldsymbol{\rho}_{(2n)}(t)$ be any solution which satisfies the master equation (and is supposedly accurate) to order $2n$, then
\begin{align}
\frac{d}{dt} \boldsymbol{\rho}_{(2n)}(t) &= \boldsymbol{\mathcal{L}}(t) \, \boldsymbol{\rho}_{(2n)}(t) + \mathcal{O}(c^{2n+2}) \, .
\end{align}
Furthermore consider the order-$2n$ state
\begin{equation}
\boldsymbol{\rho}_{(2n)}'(t) \equiv \boldsymbol{\rho}_{(2n)}(t) + \boldsymbol{\delta\!\rho}_{[2n]}(t) \, ,
\end{equation}
where $\boldsymbol{\delta\!\rho}_{[2n]}$ is an order-$2n$ traceless and diagonal (in the energy basis) perturbation for which
\begin{align}
\boldsymbol{\delta\!\rho}_{[2n]}(0) &= 0 \, , \\
\frac{d}{dt} \boldsymbol{\delta\!\rho}_{[2n]}(t) &= \mathcal{O}(c^{2n+2}) \, ,
\end{align}
so that both $\boldsymbol{\rho}_{(2n)}(t)$ and $\boldsymbol{\rho}_{(2n)}'(t)$ share the same initial conditions,
and the discrepancy $\boldsymbol{\delta\!\rho}_{[2n]}(t)$ grows slowly with the perturbation as to also satisfy
\begin{align}
\frac{d}{dt} \boldsymbol{\rho}'_{(2n)}(t) &= \boldsymbol{\mathcal{L}}(t) \, \boldsymbol{\rho}'_{(2n)}(t) + \mathcal{O}(c^{2n+2}) \, .
\end{align}
given that $\boldsymbol{\mathcal{L}}_0 \, \boldsymbol{\delta\!\rho}_{[2n]}(t) = 0$ by construction.
This demonstrates that, for non-perturbative durations of time, there is an order $2n$ ambiguity in the stationary (e.g. diagonal) entries of all solutions if one only compares terms up to order $2n$.
This proof also applies to time-nonlocal master equations, replacing perturbative contributions to the Liouvillian with corresponding memory-kernel operators.
Next we will proceed to our main proofs where we show how this issue arises, that this is the full extent of the problem, and precisely how it can be remedied.

%%%%%%%%%%%
\section{Late-time accuracy}
It is clear that if Eqs.~\eqref{eq:TCME} and \eqref{eq:PerturbExp} are well defined then for sufficiently short times an order-$2n$ master equation (in which the sum in Eq.~\eqref{eq:PerturbExp} only includes terms up to order $2n$) can produce a solution that is also accurate to order $2n$.
We find that for longer spans of time, and in particular the late-time regime wherein the master equation assumes its stationary limit \eqref{eq:Linf}, solutions to the order-$2n$ master equation are only accurate to order $2n\!-\!2$.
The reason is an ultimately mundane but slightly subtle result of degenerate perturbation theory.
In this section we will address the late-time stationary dynamics, and then in following sections we will address the full-time dynamics, including the crossover from consistent accuracy to loss of accuracy.

Assuming we have the perturbative expansion of a stationary master equation (i.e., an expansion of $\boldsymbol{\mathcal{L}}$),
we then seek perturbative solutions obtained by applying canonical perturbation theory of the eigenvalue problem \cite{QOS}
\begin{align}
\boldsymbol{\mathcal{L}} \, \mathbf{o} &= f \, \mathbf{o} \, ,
\end{align}
where $\mathbf{o}$ is a Hilbert-space eigen-operator and $f$ is its corresponding eigen-value.
We already know the zeroth-order solutions
\begin{align}
\boldsymbol{\mathcal{L}}_{[0]} \,  \ket{\omega_i} \!\! \bra{\omega_j} &= -\imath \, \omega_{ij} \ket{\omega_i} \!\! \bra{\omega_j} \, ,
\end{align}
where $\mathbf{H} \ket{\omega} = \omega  \ket{\omega}$ and $\omega_{ij} = \omega_i - \omega_j$ denote the (free) energy basis of the system.
In the appropriate regime of validity, exact solutions to the perturbative master equation should agree with the perturbative solutions to the exact master equation up to the appropriate order.
Note that perturbation theory with master equations is always degenerate perturbation theory as $\omega_{ii} = \omega_{jj} = 0$.
This inevitably-degenerate subspace corresponds to the space of operators that are diagonal in the energy basis of the free system.
For simplicity let us assume no other degeneracy in the spectrum of the free Liouvillian (though the possibility of extra degeneracy or near degeneracy arising from resonance can be suitably dealt with).

Perturbation theory tells us that the second-order corrections to all eigenvalues and eigenoperators of $\boldsymbol{\mathcal{L}}$ outside the degenerate subspace (off-diagonal operators) can be computed using only the second-order master equation:
\begin{align}
f_{ij}^{[2]} &= \bra{\omega_i} \boldsymbol{\mathcal{L}}_{[2]}\! \left\{  \ket{\omega_i}\!\! \bra{\omega_j} \right\} \ket{\omega_j} \, , \\
\bra{\omega_{i'}} \mathbf{o}_{ij}^{[2]} \ket{\omega_{j'}} &= \frac{ \bra{\omega_{i'}} \boldsymbol{\mathcal{L}}_{[2]}\! \left\{  \ket{\omega_i}\!\! \bra{\omega_j} \right\} \ket{\omega_{j'}} }{ -\imath(\omega_{ij}-\omega_{i'\!j'}) } \, . 
\end{align}
As is usual in degenerate perturbation theory,
to compute corrections to eigen-operators from the degenerate subspace, which all satisfy $\boldsymbol{\mathcal{L}}_{[0]} \, \mathbf{o}^{[0]} = \mathbf{0}$,
we must diagonalize $\boldsymbol{\mathcal{L}}$ in the degenerate subspace.
This is equivalent to finding the correct linear combination of eigen-operators which branch under perturbation.
The associated characteristic equation can be written
\begin{align}
\mathbf{W} \, \vec{\mathbf{o}} &= f \, \vec{\mathbf{o}} \, , \label{eq:Pauli} \\
[\![\vec{\mathbf{o}}]\!]_i &\equiv \bra{\omega_i} \mathbf{o} \ket{\omega_i} \, ,
\end{align}
where $\vec{\mathbf{o}}$ denotes the degenerate-subspace projection of $\mathbf{o}$ represented as a vector, i.e. diagonal entries of the eigen-operator while in the free energy basis,
and $\mathbf{W}$ defines the Pauli master equation
\begin{equation}
[\![\mathbf{W}]\!]_{ij} = \bra{\omega_i} \boldsymbol{\mathcal{L}}\! \left\{  \ket{\omega_j}\!\! \bra{\omega_j} \right\} \ket{\omega_i} \, ,
\end{equation}
which is the degenerate-subspace projection of $\boldsymbol{\mathcal{L}}$ represented as a matrix, i.e. master-equation super-operators which map diagonal entries to diagonal entries.
Therefore Eq.~\eqref{eq:Pauli} must be solved for with $\mathbf{W}_{\![2]}$ exactly,
and then the further effects of $\mathbf{W}_{\![4]}$, $\mathbf{W}_{\![6]}$, etc., can be incorporated via canonical perturbation theory.
[Note that this is slightly more complicated than the usual canonical perturbation in the Schr\"{o}dinger equation where one knows the Hamiltonian perturbation exactly.]
The eigenvalues obtained in diagonalizing $\mathbf{W}_{\![2]}$ give the second-order corrections $f^{[2]}$ to the eigenvalues of $\boldsymbol{\mathcal{L}}$ and the correct zeroth-order eigenoperators $\mathbf{o}^{[0]}$ for the degenerate subspace.
Degenerate perturbation theory tells us that,
in order to calculate each $\vec{\mathbf{o}}_i^{[2]}$ for the degenerate subspace,
one actually requires $\mathbf{W}_{\![4]}$ from the fourth-order master equation; it will contribute the second-order correction
\begin{equation}
\sum_{j \neq i} \frac{ \left( \vec{\mathbf{o}}_j^{[0]} \right)^{\!\!\star} \mathbf{W}_{\![4]} \left( \vec{\mathbf{o}}_i^{[0]} \right) }{f_i^{[2]}-f_j^{[2]}} \vec{\mathbf{o}}_j^{[0]} \, ,
\end{equation}
where $\vec{\mathbf{o}}_i^\star$ is the left eigen-vector of $\mathbf{W}$ such that $\vec{\mathbf{o}}_i^\star \, \mathbf{W} = \vec{\mathbf{o}}_i^\star \, f_i$ and $ \vec{\mathbf{o}}_j^\star \,  \vec{\mathbf{o}}_i = \delta_{ij}$.
Such corrections would be fourth order in a non-degenerate problem, but because the free Liouvillian is always degenerate,
they become second order as the relevant lowest-order nonvanishing eigenvalue splitting is always second order here.
Without this information from the fourth-order master equation, one cannot generate the complete second-order solution.

Finally note that this requirement must extend even to exact solutions of the perturbative master equation.
A perturbative solution to the second-order master equation will be equivalent to solving the full master equation perturbatively and then artificially setting $\boldsymbol{\mathcal{L}}_{[4]}$ and all higher-order contributions to the Liouvillian to vanish.
From this and the preceding perturbative analysis 
we know that the second-order perturbative solutions to the exact and second-order master equations must differ by a term that is $\mathcal{O}(c^{2})$.
Since the exact solutions to each given master equation must differ from the corresponding second-order perturbative solutions by terms of $\mathcal{O}(c^{4})$,
we can conclude from our analysis that even the exact solution to the second-order master equation differs from the exact solution to the full master equation by a term of $\mathcal{O}(c^{2})$.
In the final section we use the example of quantum Brownian motion, where an exact solution is available, to show that the second-order corrections arising from the fourth-order Liouvillian are indeed present.

More generally, the same argument tells us that while the short-time accuracy of an order-$2n$ master equation can also be order $2n$,
the long-time accuracy can only be order $2n\!-\!2$.
To obtain order-$2n$ solutions one requires not only the order-$2n$ master equation but in addition the order-$\left(2n\!+\!2\right)$ Pauli master equation.
In particular, the second-order master equation after taking the \emph{rotating-wave approximation} \cite{Breuer02}
will contain just enough terms to generate solutions which are accurate to zeroth-order \cite{RWA}.
The full second-order master equation improves upon this but not enough to generate the full second-order solutions.

Among the information missing due to the second-order errors of the solution to the second-order master equation are important contributions to the asymptotic state of the system.
When coupled to a thermal reservoir the system must asymptote to $\boldsymbol{\rho} \propto e^{-\beta \, \mathbf{H}}$ for vanishing system-environment coupling (though this may also happen in other, very specific approximations \cite{Geva00}).
One often desires to find the additional environmentally induced system-system correlations (and possibly entanglement) provided by perturbative corrections, but these will not be given correctly by directly finding the steady state of the second-order master equation.
However, at least for zero-temperature noise, it is still possible to easily construct via other methods the order-$2n$ corrections using only order-$2n$ master equation coefficients and limits thereof \cite{QOS,Dipole}.

Another important characteristic that is mangled by the second-order master equation is positivity, as was mentioned in the introduction.
The second-order inaccuracies that arise from using the second-order master equation imply that the diagonal elements of the density matrix in the (free) energy basis are off by second-order terms.
This can lead to second-order violations of positivity.
In fact, this is almost guaranteed at low temperature, where any off-diagonal perturbation to the ground state will immediately cause second-order positivity violation,
given that the necessary inequality
\begin{equation}
\rho_{ii} \, \rho_{jj} \geq \rho_{ij} \, \rho_{ji} \, ,
\end{equation}
cannot be satisfied with the left-hand side vanishing at zeroth-order and not perturbed to the correct second-order values.

%%%%%%
\section{Full-time accuracy}
In analyzing the full-time accuracy of time-dependent master equations, first we will show that the short-time solutions are accurate to the order of the master equation,
and then we will show that longer-time solutions display accuracy loss.
The timescale for this transition is determined by the frequency perturbations, e.g. $1/f_{[2]}$.

To analyze the short-time behavior we rotate to the interaction picture defined
\begin{align}
\underline{\boldsymbol{\rho}}(t) & \equiv \mathbf{G}_0^{-1}(t) \, \boldsymbol{\rho}(t) \, , \\
\mathbf{G}_0(t) \, \boldsymbol{\rho} & \equiv e^{-\imath \mathbf{H} t} \, \boldsymbol{\rho} \, e^{+\imath \mathbf{H} t} \, ,
\end{align}
wherein the master equation is now given by
\begin{align}
\frac{d}{dt} \underline{\boldsymbol{\rho}}(t) &= \underline{\boldsymbol{\delta\!\mathcal{L}}}(t) \, \underline{\boldsymbol{\rho}}(t) \, , \\
\underline{\boldsymbol{\delta\!\mathcal{L}}}(t) & \equiv \mathbf{G}_0^{-1}(t) \, \boldsymbol{\delta\!\mathcal{L}}(t) \, \mathbf{G}_0(t) \, , \\
\boldsymbol{\delta\!\mathcal{L}}(t) &\equiv \boldsymbol{\mathcal{L}}(t) -\boldsymbol{\mathcal{L}}_0 \, ,
\end{align}
and so the interaction-picture dynamics are strictly perturbative.
Short-time solutions can be obtained from the Neumann series
\begin{align}
\underline{\boldsymbol{\rho}}(t) &= \underline{\mathbf{G}}(t) \, \boldsymbol{\rho}(0) \, , \\
\underline{\mathbf{G}}(t) &= \mathbf{1} + \int_0^t \!\! d\tau \, \underline{\boldsymbol{\delta\!\mathcal{L}}}(\tau) + \int_0^t \!\! d\tau \! \int_0^\tau \!\! d\tau' \, \underline{\boldsymbol{\delta\!\mathcal{L}}}(\tau) \, \underline{\boldsymbol{\delta\!\mathcal{L}}}(\tau') + \cdots \, ,
\end{align}
where the order-$2n$ solution is fully determined by $\mathcal{L}_{[2n]}(t)$.
However, such solutions are inherently secular in time.
If $f_{[2]}$ denotes the second-order frequency perturbations, e.g. dissipation and diffusion rates, then the above solutions (at second order) are only good for times $t \ll 1/f_{[2]}$.
This is the regime wherein perturbative master equations are ensured to provide matching accuracy in their solutions.

For longer spans of time, one must resort to time-ordered integration for solutions.
For weak coupling the master equation can asymptote to its stationary value within timescales much shorter than $1/f_{[2]}$,
and so one can apply the stationary master equation and our corresponding proof of accuracy loss.
More generally one may consider the behavior of the time-dependent eigen-value equation
\begin{align}
\boldsymbol{\mathcal{L}}(t) \, \mathbf{o}(t) &= f(t) \, \mathbf{o}(t) \, ,
\end{align}
so that the time-translation generator may be given by its spectral decomposition
\begin{equation}
\boldsymbol{\mathcal{L}}(t) = \sum_k f_k(t) \, \mathbf{o}_k(t) \, \mathbf{o}^\star_k(t) \, .
\end{equation}
Again, the order-$2n$ master equation can only determine the perturbatively-stationary eigen-operators $\mathbf{o}(t)$ to within order $2n\!-\!2$.
Given that the time-dependent basis of the time-translation generator cannot be determined to second order, neither can the solutions.

One might be concerned with how the proof of short-time accuracy is compatible with this proof of full-time accuracy loss.
In fact, the short-accuracy occurs within a span of time $0<t \ll 1/f_{[2]}$, which is not sufficient enough to accumulate full-order contributions from the perturbation.
Therefore the regime of short-time accuracy is a rather trivial result.

%%%%%%%%%
\section{Time non-local accuracy}
Corresponding to the time-local master equation \eqref{eq:TCME} is the time-nonlocal master equation
\begin{align}
\frac{d}{dt} \boldsymbol{\rho}(t) &= \int_0^t \!\! d\tau \, \boldsymbol{\mathcal{K}}(t\!-\!\tau) \, \boldsymbol{\rho}(\tau) \, ,
\end{align}
first derived via the projection-operator formalism of Nakajima \cite{Nakajima58} and Zwanzig \cite{Zwanzig60}.
The two representations are contrasted in Ref.~\cite{Yan05,Chruscinski10,QOS}.
The nonlocal kernel $\boldsymbol{\mathcal{K}}(t)$ also has a perturbative expansion with zeroth-order dynamics given by
\begin{equation}
\boldsymbol{\mathcal{K}}_{[0]}(t) =2 \, \delta(t) \, \boldsymbol{\mathcal{L}}_{[0]} \, .
\end{equation}
which is time-local and unitary.
Solutions are most easily calculated in the Laplace domain wherein one has the kernel
\begin{align}
\hat{\boldsymbol{\mathcal{K}}}(s) &= \int_0^\infty \!\! dt \, e^{-ts} \, \boldsymbol{\mathcal{K}}(t) \, , \\
\hat{\boldsymbol{\mathcal{K}}}_{[0]}(s) &= \boldsymbol{\mathcal{L}}_{[0]} \, . \label{eq:K0}
\end{align}
Perturbative solutions can then be acquired by solving the nonlocal eigen-value equation \cite{QOS}
\begin{equation}
\hat{\boldsymbol{\mathcal{K}}}(s) \, \hat{\mathbf{o}}(s) = \hat{k}(s) \, \hat{\mathbf{o}}(s) \, ,
\end{equation}
where from Eq.~\eqref{eq:K0} the nonlocal eigen-system must be a perturbation of the free system-energy eigen-system,
and therefore our proof of accuracy loss will carry over.
The order-$2n$ master equation can only determine the perturbatively-stationary eigen-operators $\hat{\mathbf{o}}(s)$ to within order $2n\!-\!2$.

%%%%%%%%%%%
\section{Example: QBM} \label{sec:QBM}
As an example of an exactly-solvable open system, let us consider the master equation of an oscillator bilinearly coupled (position-position) to an environment of oscillators initially in a thermal state \cite{HPZ92}:
\begin{widetext}
\begin{align}
\frac{d}{dt} \boldsymbol{\rho} = \left[ -\imath \, \mathbf{H}_\mathrm{R} , \boldsymbol{\rho} \right] - \imath \, \Gamma \left[ \mathbf{x} , \left\{ \mathbf{p} , \boldsymbol{\rho} \right\} \right] - M D_{pp} \left[ \mathbf{x} , \left[ \mathbf{x} , \boldsymbol{\rho} \right] \right] - D_{xp} \left[ \mathbf{x} , \left[ \mathbf{p} , \boldsymbol{\rho} \right] \right] \, ,
\end{align}
\end{widetext}
where $\mathbf{H}_\mathrm{R}$ is the system Hamiltonian but with frequency $\Omega_\mathrm{R}$, %I don't want to say renormalized, really it is effective
$\Gamma$ is the dissipation coefficient,
$D_{pp}$ and $D_{xp}$ are the regular and anomalous diffusion coefficients.
This master equation describes the dynamics of damped nano-mechanical resonators at low temperature, among other physical systems.

In Ref.~\cite{QBM} exact solutions are given with full time dependence, and it is from this reference that we take all of the following results.
Let us consider Ohmic coupling to the bath with damping kernel $\gamma(t) = 2 \, \gamma_0 \, \delta_\Lambda(t)$,
where $\delta_{\Lambda}(t)$ is a representation of the delta function in the high-frequency cutoff limit $\Lambda \to \infty$.
[The damping kernel, and thus $\gamma_0$, is second order in the system-environment interaction $c$.]
The homogeneous coefficients quickly asymptote to $\Omega_\mathrm{R} = \Omega$ and $\Gamma = \gamma_0$ within the cutoff timescale,
whereas the diffusion coefficients asymptote to
\begin{align}
D_{\!xp} &= + \gamma_0 \mbox{Im}\!\left[ \mathcal{I}_0 \right] \, , \label{eq:lateDxp} \\ 
D_{\!pp} &=  2 \gamma_0 T + \gamma_0 \mbox{Im}\!\left[ \! \left( \gamma_0 + \imath \tilde{\Omega} \right)\! \mathcal{I}_0 \right] \label{eq:lateDpp} \, , \\
\mathcal{I}_0 &\equiv \frac{2}{\pi} \left(\imath + \frac{\gamma_0}{\tilde{\Omega}}\right) \left\{ \!\mbox{H}\!\left( \frac{\Lambda}{2 \pi T} \right) \!-\! \mbox{H}\!\left( \!\frac{\gamma_0 \!+\! \imath \tilde{\Omega} }{2 \pi T}\! \right) \right\} \, , \\
\tilde{\Omega} &\equiv \sqrt{\Omega^2 - \gamma_0^2} \, ,
\end{align}
mostly within the system timescale, but also hastened by temperature.
In all coefficients we have neglected terms of order $\mathcal{O}(1/\Lambda)$.
H here is the harmonic number function, which is asymptotically logarithmic and yet H$(0)=0$.
Therefore both diffusion coefficients contain logarithmic cutoff sensitivities,
though the sensitivity is present in the anomalous diffusion coefficient at second order,
whereas it does not appear in the regular diffusion coefficient until fourth order.

In the stationary limit, the system relaxes into a Gaussian state with phase-space covariance
\begin{align}
\boldsymbol{\sigma}_T &= \left[ \begin{array}{cc} \frac{1}{M \Omega_\mathrm{R}^2} \left( \frac{1}{2\Gamma} D_{pp} - D_{xp} \right) & 0 \\ 0 & \frac{M}{2\Gamma} D_{pp} \end{array} \right] \, .
\end{align}
One can see that for a second-order master equation,
the contribution from the regular diffusion $D_{pp} / \Gamma$ starts at zeroth order,
while the contribution from anomalous diffusion $D_{xp}$ starts at second order.
The full second-order contribution from the regular diffusion actually requires the fourth-order coefficients.

In the exact calculation, or in any consistent perturbative calculation, the logarithmic cutoff sensitivities present in the diffusion coefficients actually cancel in the position uncertainty.
In this sense the anomalous diffusion coefficient acts as an anti-diffusion coefficient and this behavior will also occur for supra-Ohmic couplings.
If one were to naively apply the second-order diffusion coefficients, and solve the master equation exactly, then one would obtain a mixed-order result and the logarithmic cutoff sensitivities would not precisely cancel.
The position uncertainty would contain a second-order \emph{negative} $\log(\Lambda)$ contribution.
For sufficiently large cutoff frequencies, the Heisenberg uncertainty principle would be violated.
For even larger frequencies, the covariance would become negative.
In any case the second-order master equation would produce a (supposedly) second-order position uncertainty which is an underestimation of the true second-order uncertainty.

%%%%%%%%%%%
\section{Discussion}
We have shown that even when provided with a stationary master equation describing dynamics that are amenable to perturbative solution, 
the solutions to an order-$2n$ perturbative master equation are, in general, only accurate to order-$\left(2n\!-\!2\right)$, a step down from that of the master equation itself.
This has a wide-range of implications upon the common use of second-order master equations and related master equations derived from second-order dynamics:
the Redfield, Born-Markov, and many Lindblad equations.
Moreover, not even a nonlocal representation, such as with the Nakajima-Zwanzig master equation can avoid this effect.
This is to be expected as a thorough analysis of time-local and nonlocal dynamics shows their asymptotics to be perturbatively the same \cite{QOS}.

To be more specific, the second-order master equation can provide all second-order timescales and off-diagonal density matrix elements (in the free energy basis).
However it can only provide the diagonal matrix elements with zeroth-order accuracy, and the missing information is the most relevant to positivity in the low-temperature regime.
Therefore the second-order master equation can produce second-order positivity violations, whereas the full second-order solutions are positive to second-order.
Likewise, the steady state of the second-order master equation may only agree with the steady state of the full master equation to zeroth order.  
More generally, the predicted expectation of observables will typically be off by a second-order amount, and this certainly includes the energy or other quantities that were conserved at zeroth order.
We have shown that this inaccuracy manifests itself in the case of quantum Brownian motion through an underestimation of the position undertainty stationary limit.  The same issue also affects the predicted dynamics for a collection of atoms interacting with a shared field \cite{Dipole}, where the complete second-order solution is required make correct predictions about the sudden death of entanglement.

There are three mathematical limits in which the second-order master equation will give solutions accurate to second order:
The first is early times, where $t$ is small compared to any of the second-order damping time scales.
The second is the Markovian limit, because in this limit the second-order master equation is exact.
The third is the limit employed by Davies \cite{Davies74} where one rewrites the master equation in terms of the rescaled time parameter $\tau = c^2 t$ and then takes the limit $c \to 0$ (for $\tau \ne 0$ this effectively amounts to taking a simultaneous $t \to \infty$ limit).
In this limit all corrections to the eigenoperators of the Liouvillian vanish, and the only effect of the environment is to introduce damping rates through corrections to the eigenvalues, which are correctly captured by the perturbative master equation.
Thus, the inaccuracies of second-order master equation we have addressed may be sufficiently suppressed even at late times if a physical system is sufficiently close to being described by one of these limits.
Therefore our results should be most heeded in the non-Markovian regime of low temperature or long-ranged correlations in contexts where $\mathcal{O}\!\left( c^2 \right)$ discrepancies are not negligible.

\bibliography{bib}{}
\bibliographystyle{apsrev4-1}

\end{document}